# Launch Vehicle High-Energy Performance Dataset

Athul Pradeepkumar Girija [1***]

*[1]School of Aeronautics and Astronautics, Purdue University, West Lafayette, IN 47907, USA*

## ABSTRACT

The choice of the launch vehicle is an important consideration during the preliminary planning of interplanetary missions. The launch vehicle must be highly reliable, capable of imparting sufficient energy to the spacecraft to inject it on to an Earth-escape trajectory, and must fit within the cost constraints of the mission. Over the recent past, the most commonly used launchers for interplanetary missions include the Atlas V401, Atlas V551, Delta IVH, and Falcon Heavy expendable version. The NASA Launch Vehicle Performance website maintains a tool to help mission planners evaluate various launch vehicles during mission studies. However, there is no comprehensive dataset which can be used to quickly compare the launch performance and launch cost of various options. The present study compiles a dataset of the high energy performance of existing and planned launchers from open-source data and performs a quantitative comparison of the launch performance and the launch cost per kg. The Falcon Heavy expendable offers the lowest cost-per-kg for high-energy launches, with only $0.075M per kg. The Vulcan Centaur offers comparable performance to the Falcon Heavy. The results indicate Falcon Heavy Expendable and the Vulcan Centaur will be the likely choice for several future missions.

---

***** To whom correspondence should be addressed, E-mail: athulpg007@gmail.com

## I. INTRODUCTION

The choice of the launch vehicle is an important consideration during the preliminary planning of interplanetary missions [1]. The launch vehicle must be highly reliable, capable of imparting sufficient energy to the spacecraft to inject it on an Earth-escape trajectory to begin its interplanetary cruise, and must fit within the cost constraints of the proposed mission or program [2]. Over the recent past, the most commonly used launchers for interplanetary missions include the Atlas V401, Atlas V551, Delta IVH, and Falcon Heavy expendable version. The Psyche mission to the iron-nickel rich asteroid will be the first interplanetary science mission to fly on the Falcon Heavy [3]. The Vulcan Centaur under development is indented to replace the Atlas V and the Delta IV, and is expected to offer comparable performance to the Falcon Heavy [4]. The Space Launch System (SLS) developed by NASA is a highly-capable, but also expensive system with estimates ranging in excess of $2B per launch [5]. The NASA Launch Services Program maintains a website to help mission planners evaluate various launch vehicles during mission studies [6]. However, there is no comprehensive dataset which can be used to quickly compare the launch performance and launch cost of various options for interplanetary missions. The present study compiles a dataset of the high energy performance of existing and proposed launchers from open-source data and performs a quantitative comparison of their launch mass capability and the launch cost per kg for interplanetary missions [7].

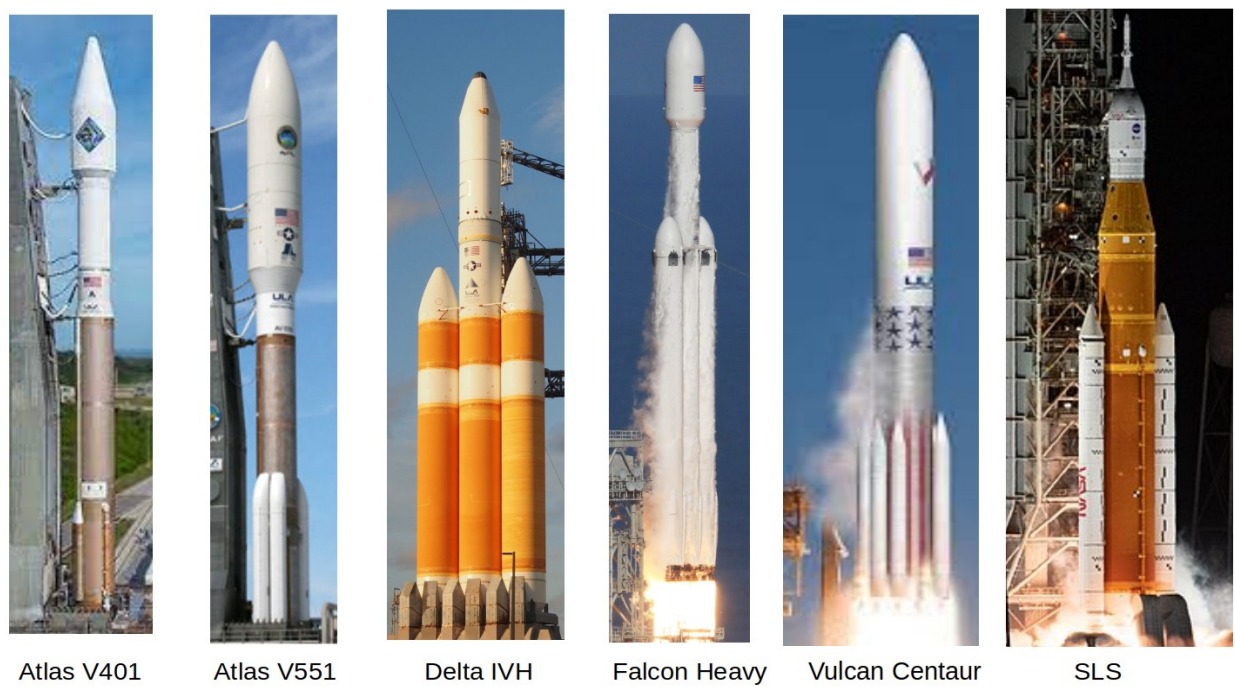

| Atlas V401 | Atlas V551 | Delta IVH | Falcon Heavy | Vulcan Centaur | SLS |

Figure 1.  Set of launchers used for interplanetary missions from the dataset.

## II. HIGH-ENERGY LAUNCH PERFORMANCE

The high energy launch performance is characterized by the mass capability at a specified C3, which is the square of the hyperbolic excess speed relative to Earth. Missions to Mars or Venus typically have launch C3 in the range of 10-20 km$^2$/s$^2$ [8], while missions to the outer planets and Kuiper Belt Objects can have much higher C3 [9]. The New Horizons mission holds the current record for the highest C3, at 157 km$^2$/s$^2$. The Parker Solar Probe was the next highest at about 154 km$^2$/s$^2$. Figure 2 shows the launch performance as a function of the C3 for the vehicles in the dataset. The Atlas V series have the relatively low high-energy performance in the dataset, but is also the least expensive. The Delta IVH offers more high-energy performance, but is also significantly more expensive. The Falcon Heavy Recoverable has poor high-energy performance, while the expendable version with the STAR48 kick stage offers excellent performance within a cost of only $150M. The Vulcan Centaur performance and cost is comparable to the Falcon Heavy. Falcon Heavy expendable is used for the Psyche mission, and is also baselined for the Uranus Orbiter and Probe Flagship mission. Even though aerocapture was not considered [10, 11], recent studies have shown that it offers considerable benefits for ice giant missions [12, 13]. Falcon Heavy expendable is also a viable option for an aerocapture missions to Uranus which offers shorter flight times [14]. SLS offers the highest launch performance, especially at very high C3 such as those for fast outer planet missions, but is estimated to cost in excess of $2B [15].

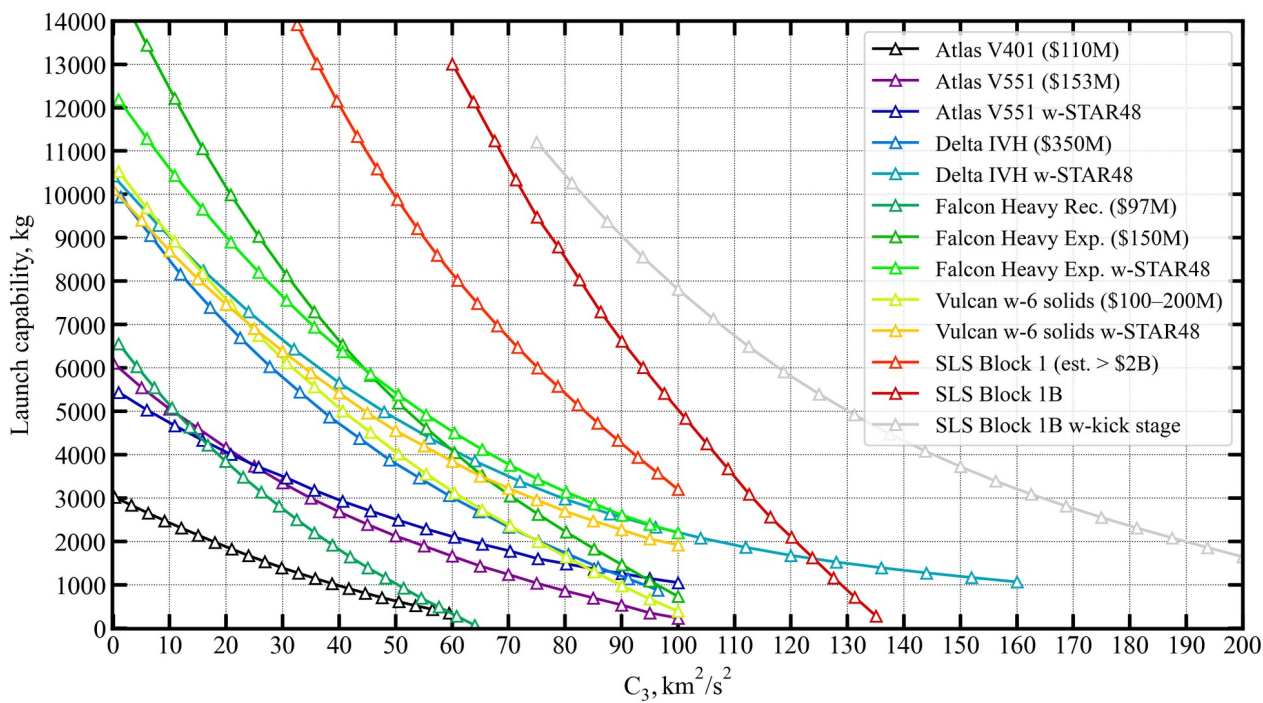

Figure 2. High-energy launch performance of the vehicles in the dataset.

## III. LAUNCHER COST COMPARISON

Figure 3 shows the estimated launcher cost based on open source data. The Falcon Heavy (FH) expendable and the Vulcan both cost approximately $150M, while offering much better performance than Atlas V551 which also costs the same. SLS is an outlier in terms of the cost compared to other vehicles in the dataset. Figure 4 shows the cost-per-kg ($M) as a function of C3. The FH expendable has the lowest cost-per-kg for high-energy launches, with only $0.075M per kg. The Vulcan Centaur offers comparable cost-per-kg to the FH expendable. These results indicate the Falcon Heavy Expendable, or the Vulcan Centaur will be the likely choice for several future missions [16, 17].

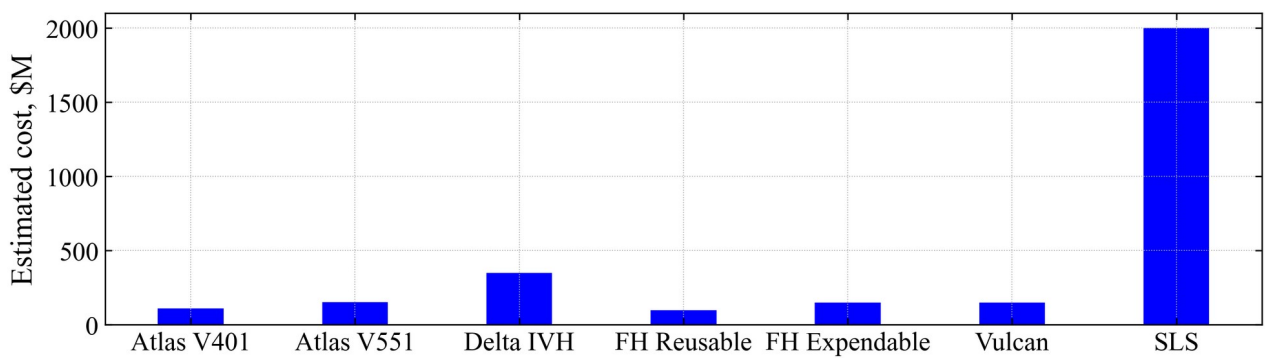

Figure 3. Cost comparison of vehicles in the dataset.

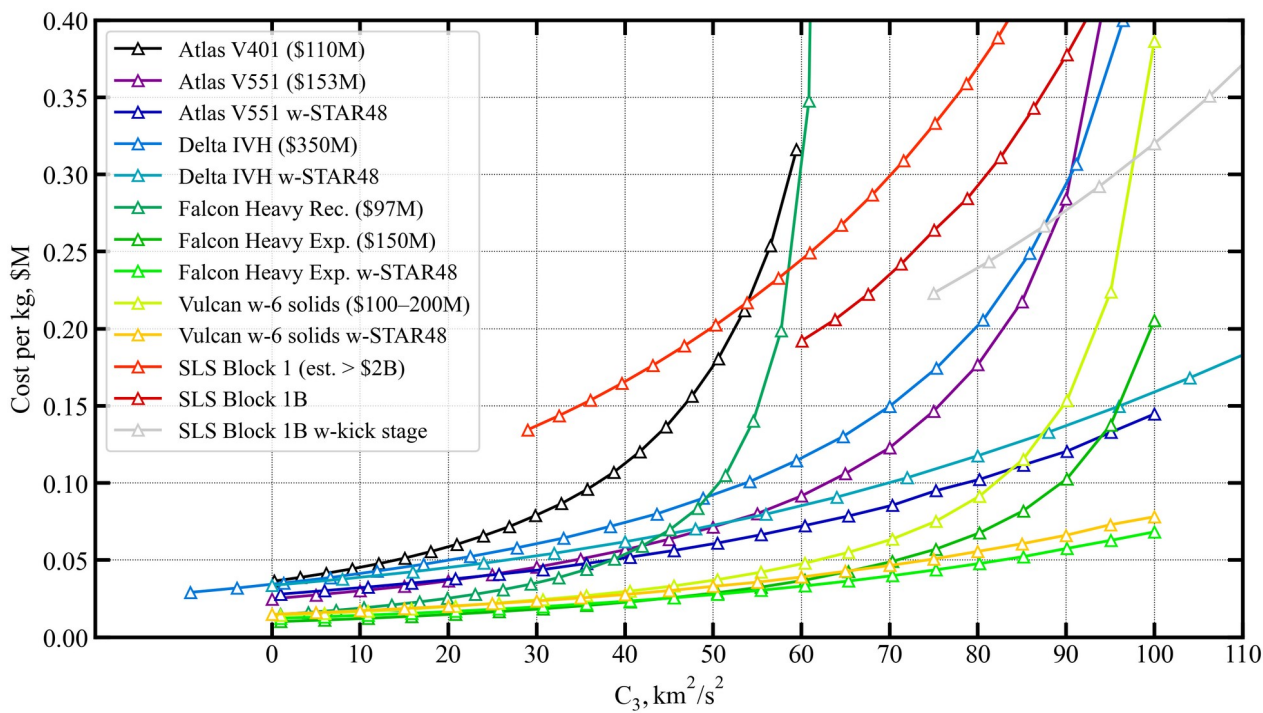

Figure 4. Cost-per-kg ($M) comparison of vehicles as a function of C3.

## IV. CONCLUSIONS

The present study compiled a dataset of the high energy performance of various launchers from open-source data. The dataset is used to performs a quantitative comparison of launch mass capability and the launch cost per kg. The The Falcon Heavy expendable offers the lowest cost-per-kg for high-energy launches, with only $0.075M per kg. The Vulcan Centaur offers comparable cost-per-kg to the FH expendable. These results indicate the Falcon Heavy Expendable, or the Vulcan Centaur will be the likely choice for several future interplanetary missions.

### DATA AVAILABILITY

The launch performance dataset used in the study is available at https://dx.doi.org/10.13140/RG.2.2.33904.02564.

The results presented in the paper were produced using the open-source Aerocapture Mission Analysis Tool (AMAT) v2.2.22. The code used to make the study results will be made available by the author upon request.